\documentclass[reprint, superscriptaddress, english, preprintnumbers, 
amsmath,amssymb,aps,prx,floatfix]{revtex4-2}
\usepackage{graphicx}
\usepackage{dsfont}
\usepackage{xcolor}
\usepackage{multirow}
\usepackage[normalem]{ulem}
\usepackage{enumitem} 
\usepackage{lineno}
\usepackage{bm}
\usepackage{charter} 
\definecolor{refcol}{RGB}{34,34,178}
\usepackage[colorlinks,linkcolor=blue,citecolor=blue,urlcolor=blue]{hyperref}
\usepackage{microtype}
\usepackage{float}
\usepackage[caption = false]{subfig}
\usepackage{appendix}
\usepackage{multirow}
\usepackage{soul}

\graphicspath{{./Fig/}}
\begin{document}

\title{
Chemical freeze-out parametrization with mean field repulsive hadron resonance gas model 
}      
\author{Sunny Kumar Singh}
\email{sunny.singh@iitgn.ac.in}
 \affiliation{Indian Institute of Technology Gandhinagar, Palaj, 
 Gujarat 382355}
 
\author{Nachiketa Sarkar}
\email{nachiketa.sarkar@gmail.com}
\affiliation{School of Physical Sciences, National Institute of Science 
Education and Research, An OCC of Homi Bhabha National Institute, 
Jatni-752050, India} 

\author{Deeptak Biswas}
\email{deeptakb@gmail.com}
\affiliation{The Institute of Mathematical Sciences, a CI of Homi 
Bhabha National Institute, Chennai, 600113, India} 

\date{\today}
\begin{abstract}
We have examined the chemical freeze-out surface of the heavy-ion 
collision experiments within an interacting hadron resonance gas model. 
By considering repulsive interaction among hadrons in the mean-field 
level, we have suitably parameterized the freeze-out surface by fitting 
the yield data of mid-rapidity for the most central collision, for the 
collision energy available in AGS, RHIC (BES), and LHC programs. To 
suitably account for the repulsive interaction among mesons and (anti-) 
baryons, we have introduced phenomenological parameters $K_M$ and $K_B$ 
in the freeze-out parametrization. Although a finite value of these two 
parameters seem to be necessary to have an improved normalized 
\emph{chi-square}, the effect on the rest of the parameters like 
temperature and relevant chemical potentials seem to be within the 
standard variance.
\end{abstract}
\maketitle
\section{Introduction}
The investigation of the phase structure of strongly-interacting matter 
stands as a pivotal and fundamental inquiry within the realm of 
ultra-relativistic heavy-ion physics. To comprehend the particle spectra 
observed in these experiments, statistical thermal models inspired by 
quantum chromodynamics (QCD) are employed. In particular, the transverse 
momentum ($p_T$) integrated rapidity spectra (namely $dN/dY$) are frozen 
onward the chemical freeze-out (CFO) boundary and helps to map the 
freeze-out surface on the phase diagram via the CFO parametrization with 
temperature ($T$) and baryon chemical potentials ($\mu_B$) 
\cite{Stock2020}. For the past few decades, the Hadron resonance Gas 
(HRG) model has been successfully describing the abundance of hadrons in 
collisions across a wide range of energies, from the SchwerIonen-
Synchrotron (SIS) to the Large Hadron Collider (LHC) 
\cite{Cleymans:1992zc, Cleymans:1998fq, Cleymans:1999st, 
Becattini:2000jw, Andronic:2005yp, Andronic:2008gu}. The success of the 
HRG model, coupled with the lack of reliable first-principle theories 
that can provide such parameterization for both high and low baryon 
density regions of the phase diagram have firmly established HRG as one 
of the most widely utilized models in this field.

The simplest version of the HRG model is the ideal HRG model (IHRG) 
\cite{Cleymans:1999st, Becattini:2000jw, Andronic:2008gu}, where 
attractive interactions among hadrons in a dilute hadron gas can be 
approximated by treating higher mass resonances as stable particles.  
Initially proposed within the relativistic virial expansion framework, 
using the $S$-matrix approach \cite{Dashen:1969ep}, this model allows for 
the calculation of various thermodynamic quantities 
\cite{Prakash:1993bt}. However, the IHRG model encountered discrepancies 
in different thermodynamic quantities when compared to lattice QCD 
results \cite{Bellwied:2015lba, Bazavov:2013dta}, particularly at the 
temperature range above the pseudo-critical value. Additionally, an 
excess in the pion number density at chemical freeze-out was 
observed\cite{Yen:1997rv}, indicating the need to incorporate short-range 
repulsive interactions between hadrons to achieve more accurate Equations 
of State (EoS) and realistic estimations of the chemical freeze-out 
boundary.

One of the frequently employed methods to model the short-range repulsion 
is the Excluded Volume Hadron Resonance Gas (EVHRG) model~ 
\cite{Andronic:2012ut, Rischke:1991ke, Hagedorn:1980kb, Begun:2012rf, 
Yen:1997rv, Tiwari:2011km}. In this model, repulsive interactions are 
taken into account by incorporating an impenetrable volume surrounding 
the individual hadrons. Several versions of the EVHRG model have been 
proposed in the literature to determine the strength of short-range 
repulsive interactions through comparisons with lattice QCD calculations 
or experimental data. These include the diagonal EVHRG model 
\cite{Yen:1997rv}, the cross-terms EVHRG \cite{Gorenstein:1999ce, 
Vovchenko:2016ebv}, the mass-dependent EVHRG 
\cite{Noronha-Hostler:2009iqu, Vovchenko:2015cbk}, and the Flavor-
dependent EVHRG model \cite{Alba:2016hwx}. Another phenomenological 
approach to include the interaction is the Van der Waals Hadron Resonance 
Gas (vdWHRG) model, which explicitly incorporates both repulsive and 
attractive interactions between baryons and anti-baryons 
~\cite{Vovchenko:2015xja, Vovchenko:2015vxa, Samanta:2017yhh, 
Sarkar:2018mbk, Sarkar:2023cnq}.

The repulsive interactions between the various baryon-baryon and meson-
meson can also be incorporated at the mean-field level 
\cite{Olive:1980dy, Olive:1982we, Kapusta:1982qd, Huovinen:2017ogf}. 
The interacting part of the pressure is added along with the ideal one and
modification is introduced into the statistical model by shifting the 
energy of each particle by an amount equal to $U(n)=Kn$ where $n$ is the 
total hadron number density. One can incorporate the mean-field 
coefficients $K_M$ and $K_B$ to scale the repulsive interaction strength 
among the mesons and baryons respectively. Recent works 
\cite{Huovinen:2017ogf} have augmented the mean-field coefficients $K_B$ 
from lattice data of $\chi_2^B - \chi_4^B $ and $\chi_2^B - \chi_6^B$. In 
another investigation, suitable values of $K_B$ and $K_M$ were estimated 
by fitting lattice QCD data of bulk observables, cumulants, and the speed 
of sound \cite{Kadam:2019peo}.

In this study, we have focused on constraining the mean-field model at 
the chemical freeze-out boundary by comparing it with experimental yields 
through a $\chi^2$ minimization procedure. Previous applications of this 
mean-field model at freeze-out involved fixing the repulsive strength 
parameter $K$ to explain the  data of 200 $A$GeV S + Au collisions at CERN-
SPS \cite{Sollfrank:1996hd}. While earlier studies consistently suggested 
a value of $K_B = 450$ $\text{MeV fm}^{-3}$, we aim to investigate the 
collision energy dependence of these phenomenological parameters by 
analyzing the rapidity spectra. Within this approach, for the first time 
we have obtained the collision energy dependence of mean-field 
coefficients by parameterizing the chemical freeze-out surface for RHIC 
and LHC energies. 

We have organized the paper as follows. In Sec. \ref{sec:model} we give a 
short description of the ideal HRG and the MFHRG model. In Sec. 
\ref{secIII} we discuss the method we have employed to extract the 
various parameters in the model. In Sec. \ref{secIV} our results and the 
discussion of our results are provided in the context of heavy ion 
collision experiments. We conclude by giving a summary of the present 
work in Sec. \ref{sec:summary}. 
	
\section{Formalism}{\label{sec:model}}  
In the ideal hadron resonance gas model, the thermodynamic potential for each species is \cite{Hagedorn:1965st, Hagedorn:1980kb}:
\begin{equation}
\begin{aligned}
\ln &Z^{id}_i(T,\mu,V) \\
&=\pm\frac{Vg}{(2\pi)^3}\int d^3p \ln[1\pm e^{(-(E_i-\mu_i)/T)}]
\end{aligned}
\end{equation}
Where the upper(lower) sign corresponds to fermions(bosons). Here $g$ is 
the degeneracy factor and $V$ is the volume. Considering the baryon 
number ($B$), electric charge ($Q$), and strangeness ($S$), the chemical 
potential ($\mu_i$) of the $i$th hadron is determined by $\mu_i = Q_i\mu_Q 
+ S_i\mu_S + B_i\mu_B$.

The grand thermodynamic potential for the total ensemble is given by:
\begin{equation}
\ln Z^{ideal}=\sum_i\ln Z_i^{ideal}
\end{equation}
The number density of each species can be determined by:
\begin{eqnarray}
n_i &=& \frac{T}{V}\left(\frac{\partial \ln Z_i}{\partial 
\mu_i}\right)_{V, 
T} \nonumber \\
&=& \frac{g_i}{(2 \pi)^3} \int \frac{d^3 p}{\exp \left[\left(E_i-
\mu_i\right) 
/ T\right] \pm 1} .
\end{eqnarray}
One can relate the thermal abundance of the detected particles at the 
chemical freeze-out surface with the corresponding rapidity densities as 
follows:
\begin{equation}
\left.\left.\frac{d N_i}{d y}\right|_{\text {Det }} \simeq \frac{d V}{d y} 
n_i^{\mathrm{Tot}}\right|_{\text {Det }}
\end{equation}
The total number density of each species considering decays from higher 
resonances can be computed as follows:
\begin{equation}
\begin{aligned}
n_i^{Tot} =&~n_i(T,\mu_B,\mu_Q,\mu_S) \\
&+\sum_j n_j(T,\mu_B,\mu_Q,\mu_S) \times \text{Branching Ratio}
(j\rightarrow i)
\end{aligned}
\end{equation}
\subsection{Mean-Field HRG (MFHRG)}\label{sec:MFHRG}
With the inclusion of short-range repulsive interactions between hadrons 
via mean-field approach, the effective chemical potential of each particle 
species gets modified by $\mu_{\mathrm{eff},i}=\mu_i - Kn$, where \emph{K} 
is a phenomenological parameter that signifies the strength of the 
repulsive interaction and $n$ is the number density of the interacting 
species of particles \cite{Huovinen:2017ogf, Kadam:2019peo}. The pressure 
of the mean-field repulsive  model is given by:
\begin{equation}
\begin{aligned}
&P_{\mathrm{MF}}\left(T, \mu, V\right) \\
&= \pm T \sum_i \frac{g_i}{(2\pi)^3}\int d^3p \ln \left[1 \pm e^{-\left(E_i-
\mu_{\mathrm{eff},i}\right)/T}\right] + \mathcal{P}_{M, B, \bar{B}}\left(n_{M, B, 
\bar{B}}\right)
\end{aligned}
\end{equation}
Here, $\mathcal{P}$ is the factor arising from the interacting 
part, which is necessary to maintain the thermodynamic consistency 
\cite{Huovinen:2017ogf}.
\begin{equation}
\begin{aligned}
\mathcal{P}_{B\{\bar{B}\}}\left(n_{B\{\bar{B}\}}\right) & 
= \frac{1}{2} K_B n_{B\{\bar{B}\}}^2, \quad \text { (Baryons}) \\
\mathcal{P}_M\left(n_M\right) & = \frac{1}{2} K_M n_M^2, \quad \text { 
~~~~(Mesons) }
\end{aligned}
\end{equation}
The above form of interacting pressure is written considering repulsive 
interactions among meson-meson and baryon(anti-baryon)- 
baryon(anti-baryon) pairs. Here the total meson number density $n_M$ is 
calculated as:
\begin{equation}\label{Eq.nM}
\begin{aligned}
n_M=\sum_{i\in {M}}\frac{g_i}{(2 \pi)^3} \int \frac{d^3 p}{\exp 
\left[\left(E_i-\mu_{\text{eff},i}\right) / T\right] - 1} .
\end{aligned}
\end{equation} 
For mesons, $\mu_{\text{eff}, i}=\mu_i-K_Mn_M$. $K_M$ signifies the 
strength of the repulsive interactions among the meson-meson pairs. We 
have a similar equation for baryons and anti-baryon number densities:
\begin{equation}\label{Eq.nB}
\begin{aligned}
n_{B\{\bar{B}\}}=\sum_{i\in {B\{\bar{B}\}}}\frac{g_i}{(2 \pi)^3} \int \frac{d^3 p}{\exp 
\left[\left(E_i-\mu_{\text{eff},i}\right) / T\right] + 1} .
\end{aligned}
\end{equation} 
$B$ and $\bar{B}$ imply baryons and anti-baryons respectively. 
Here, the effective chemical potential of the $i$th (anti-)baryon  
$\mu_{\text{eff},i}=\mu_i-K_B n_{B\{\bar{B}\}}$. The repulsive 
interactions among the baryon-baryon and antibaryon-antibaryon pairs are 
given by the same strength parameter $K_B$. These equations are 
transcendental in nature and should be solved simultaneously with 
these two equations Eq.[\ref{Eq.nM}-\ref{Eq.nB}].

\section{Method and Data Analysis}\label{secIII}
The mid-rapidity data of hadron yields $dN/dY$ were taken from various 
experiments at 0-5\% centrality (most central) and at different energies. 
These consist of Pb-Pb collisions in LHC at a collision energy of 2760 
GeV~\cite{Abelev:2012wca, Abelev:2013xaa, ABELEV:2013zaa, 
Abelev:2013vea}. We have also included Au-Au collisions at RHIC of 200, 
130, 62.4 GeV ~\cite{Kumar:2012fb, Das:2012yq, Adler:2002uv, 
Adams:2003fy, Zhu:2012ph, Zhao:2014mva, Kumar:2014tca, Das:2014kja, 
Abelev:2008ab, Aggarwal:2010ig, Abelev:2008aa, Adcox:2002au, 
Adams:2003fy, Adler:2002xv, Adams:2006ke, Adams:2004ux, Kumar:2012np, 
Adams:2006ke}, RHIC BES of 39, 27, 19.6, 11.5, 7.7 
GeV~\cite{Adamczyk:2017iwn, STAR:2019bjj}. and in AGS at 4.85 
GeV~\cite{Ahle:1999uy, Ahle:2000wq, Klay:2003zf, Klay:2001tf, 
Back:2001ai, Blume:2011sb, Back:2000ru, Barrette:1999ry, Back:2003rw}. 
 
To extract the chemical freeze-out parameters i.e., $T$, $\mu_B$, 
$\mu_S$, $\mu_Q$ along with the parameters scaling the strength of the 
hadron-hadron interaction ($K_B$ and $K_M$), we have fitted the detected 
hadron yields with the thermal model estimations. Considering the initial 
condition of the heavy-ion collision, it is customarily practiced to fix 
$\mu_Q$ and $\mu_S$ via two constraint equations. The first constraint is 
the ratio of net baryon to net charge which remains fixed throughout the 
collision process considering the isentropic evolution 
\cite{Alba:2014eba}.
\begin{equation}
\frac{\sum_i n_i(T,\mu_B,\mu_S, \mu_Q,K_M, K_B)B_i}{\sum_i 
n_i(T,\mu_B,\mu_S,\mu_Q,K_M, K_B)Q_i}= r
\end{equation}
One can evaluate this ratio $r$ considering the number of neutrons and 
protons in the incident nuclei. For heavy nuclei like Au-Au and Pb-Pb, 
this ratio $r$ is approximately 2.5 \cite{Biswas:2020kpu}.

The conservation of strangeness along with the strangeness neutrality 
imposes another constraint:
\begin{equation}
\sum_i n_i(T,\mu_B,\mu_S, \mu_Q,K_M, K_B)S_i=0
\end{equation}
The rest of the parameters are determined by the $\chi^2$ minimization 
procedure. The $\chi^2$ is defined as:
\begin{equation}
\chi^2=\sum_i\frac{\left(
\left.\frac{d N_i}{d y}\right|_{\text {Expt}}-\left.\frac{d N_i}{d 
y}\right|_{\text {Model}}\right)^2}{\sigma_i^2} 
\end{equation}
Here, we would like to emphasize that our present analysis focuses 
exclusively on data from the most central events of the collisions, thus  
we have chosen not to incorporate the strangeness suppression factor 
$\gamma_s$ assuming a state of complete chemical equilibrium. For the 
present study, we have used data of $ \pi^\pm, K^\pm, p, \bar{p}, 
\Lambda, \bar{\Lambda}, \Xi^\pm$, as these are widely available for most 
of the collision energies. To optimize numerical efficiency and reduce 
the number of free parameters, we have fixed the $K_M$ at three different 
values, i.e. 0, 50, and 100 $\text{MeV fm}^{-3}$. In the considered HRG 
spectrum, all confirmed hadronic states up to mass 3 GeV have been 
included, with masses and branching ratios following the Particle Data 
Group~\cite{Tanabashi:2018oca}.

The statistics and systematic uncertainties in a given data have been 
added considering the quadrature method. The variance of the evaluated 
parameter set for a particular minimization procedure has been calculated 
from the $\pm 1$ deviation of the minimized $\chi^2$ per degree of 
freedom~\cite{Andronic:2005yp}. 
 \section{Result and Discussion}\label{secIV}

\begin{figure}
\hspace*{-0.6cm}
\includegraphics[height=3.2 in,width=3.8 in]{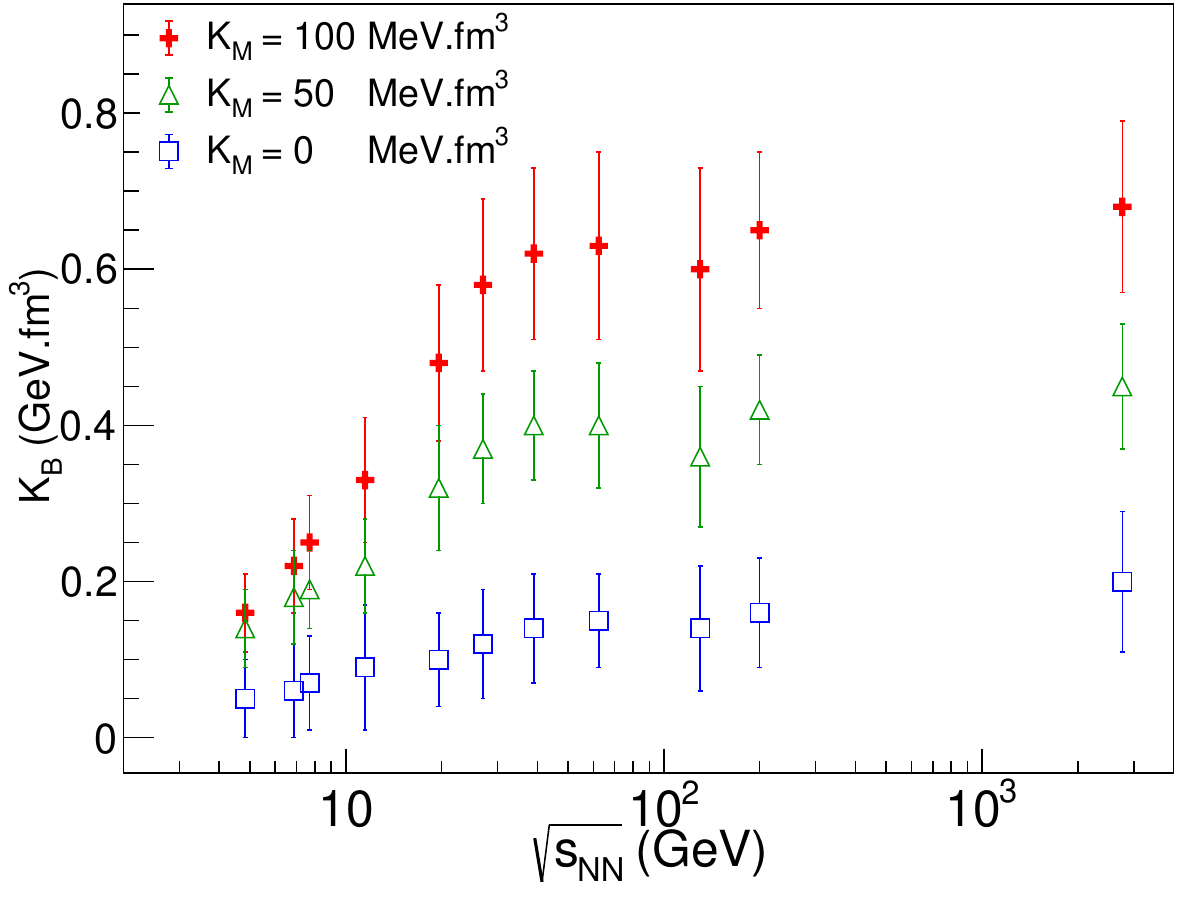}
\caption{Results from the fitted set of parameters. Blue, green, and red 
points are results for $K_M$ values of 0, 50, and 100 $\text{MeV 
fm}^{-3}$ respectively.}
\label{fig:KB}
\end{figure} 
\subsection{Variation of freeze-out parameters:}{\label{Sec:parameters}}
\begin{figure*}[htb]
\hspace*{-0.9cm}
\includegraphics[height=5.4 in,width=7.4 in]{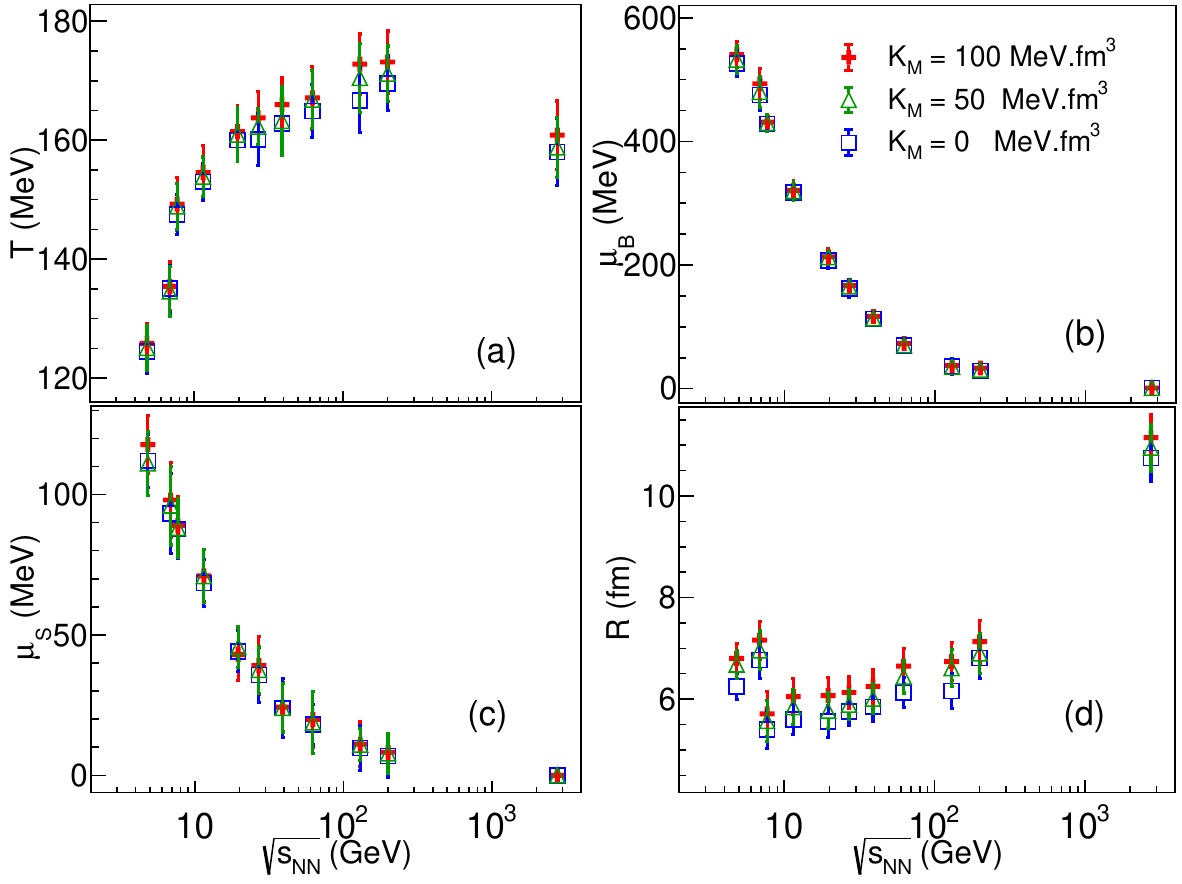}
\caption{Fitted set of parameters. Blue, green, and red points are 
results for $K_M$ values of 0, 50, and 100 $\text{MeV fm}^{-3}$ 
respectively.}
\label{fig:parameters}
\end{figure*} 
We have tabulated the fitted parameter set in Table.~\ref{tb:1}. For convenience, let us first discuss the 
variation for the mean-field coefficients, as these are the most novel 
output from our present study. Changes in other freeze-out parameters are 
commensurate with the variation in these mean-field coefficients. 
Extraction of both the $K_M$ and $K_B$ becomes numerically challenging
due to the slow convergence rate. We have fixed the values of $K_M$ to be 
0, 50, and 100 $\text{MeV fm}^{-3}$ and examined the corresponding 
values of $K_B$. For a fixed value of $K_M$, the $K_B$ increases with 
collision energies and remains similar at higher RHIC and LHC energies as 
shown in Fig.~\ref{fig:KB}. A similar saturation of thermal parameters at 
higher collision energies has been noticed earlier for temperature and 
chemical potentials \cite{Bhattacharyya:2019wag, Biswas:2020dsc}. We have 
found that even for $K_M=0.0~\text{MeV fm}^{-3}$ a non-zero value of 
$K_B$ helps achieve better $\chi^2$ per degree of freedom while fitting 
with yield data. With increasing the $K_M$ to 50 and 100 $\text{MeV 
fm}^{-3}$ the values of $K_B$ increases. In the context of heavy-ion 
collision, the total ensemble of baryons and mesons are connected via the 
constraints like net strangeness neutrality and a fixed net 
baryon-to-charge ratio. Along with these constraints, the final yield of 
mesons is predominantly influenced by the decay of various higher-mass 
baryon resonances \cite{Wheaton:2004qb}. Consequently, imposing a mean-
field repulsion in mesons necessitates a higher value of $K_B$ to 
restrict the baryon abundances, which eventually affects the final yield 
of mesons and validates the required constraints. 

Towards the lower collision energy, the medium is mainly baryon-dominated 
\cite{Cleymans:2004hj}, and the effect from the variation of $K_M$ is 
minimal. However, the system becomes meson dominated with increasing 
$\sqrt{s_{NN}}$, and the effect of $K_M$ is much more pronounced. We 
would like to emphasize that for the range of $K_M$ considered, which 
spans from 0 to 100 $\text{MeV fm}^{-3}$, the corresponding $K_B$ values 
vary between 100 and 800 (considering variances) $\text{MeV fm}^{-3}$. These 
specific values were previously explored in a hydrodynamic simulation that 
incorporated a hadronic equation of state, as mentioned in 
Ref.~\cite{Sollfrank:1996hd}. Furthermore, recent studies conducted using the 
MFHRG model have also confirmed this range of $K_B$ values as they successfully 
account for the lattice data of various charge susceptibilities 
\cite{Huovinen:2017ogf, Huovinen:2018xmq, Kadam:2019peo}.

In the top left panel of Fig.~\ref{fig:parameters}, we have shown the 
variation of freeze-out temperature with collision energy for the three 
considered values of the mesonic mean-field coefficients $K_M$ as 
mentioned earlier. For the freeze-out temperature ($T$) the difference 
from considering three different values of $K_M$ seems to be similar to 
$K_B$. The differences increase towards high collision energies, 
following the variation of $K_B$. For all three values of $K_M$, the 
temperature increases with the collision energy and becomes constant 
around $160$ MeV near higher BES energies. We want to iterate here that, 
although the qualitative behavior is similar to the usual understanding 
of our freeze-out parametrization within the ideal HRG formalism, the 
temperature value is slightly higher ($\sim$ 5 MeV) than the ideal HRG 
result. A finite value of the mean-field repulsion parameter restricts 
the number density which in turn produces a higher $T$ to fit the yields. 
\begin{figure*}
\hspace*{-0.9cm}
\includegraphics[height=5.4 in,width=7.4 in]{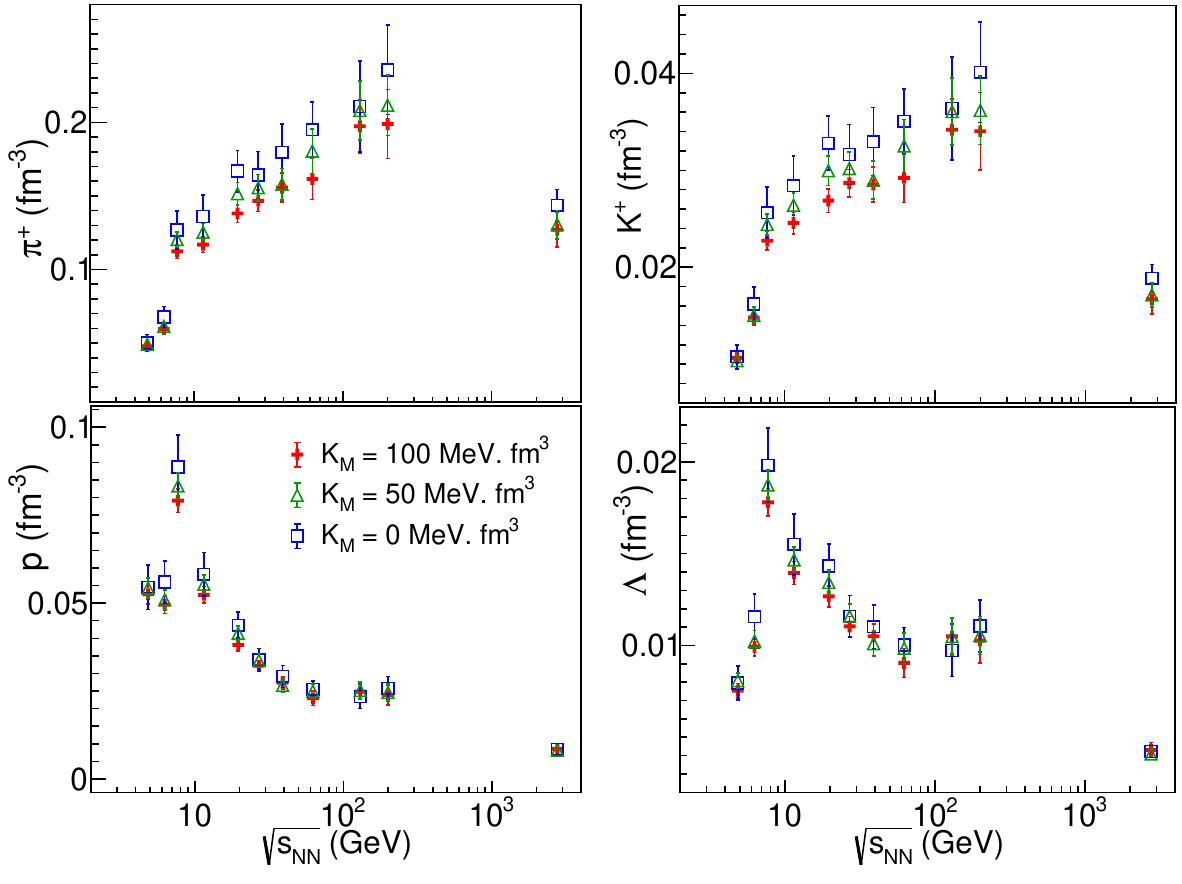}
\caption{$\sqrt{s_{NN}}$ variation of the yields of (top left) $\pi^+$, 
(top right) $K^+$, (bottom left) $p$ and (bottom right) $\Lambda$. Blue, 
green, and red points are the results for $K_M$ values of 0, 50, and 
100 $\text{MeV fm}^{-3}$ respectively. Black points are the experimental 
data.}
\label{fig:yield}
\end{figure*} 

The collision energy dependence of the baryon chemical potential is shown 
in the top right panel.  The effect of repulsion is almost negligible on the 
freeze-out values of $\mu_B$. However, at very lower collision energy, the 
effect of $K_B$ seems to induce a higher $\mu_B$ as the medium in 
dominated by baryons. The chemical potential is shifted by $K_B n_B$, so 
a higher value of $K_B$ should be accompanied by a higher $\mu_B$ to 
produce  a similar estimation of yields. The general 
behavior is similar to that of the ideal HRG parametrization. With higher 
collision energies the baryon stopping diminishes so the medium tends to 
form with lower net charges ($B, Q$, and $S$), which results in a lower 
value of chemical potentials in the freeze-out parametrization. At lower 
collision energy this behavior induces a high value of $\mu_B$, which 
tends to be zero at higher RHIC LHC energy

The strange chemical potential follows the trend of $\mu_B$. A finite 
$\mu_B$ results in the dominance of the hyperons over the anti-hyperons, 
on the other hand, the strangeness-neutrality constraint demands the 
cancellation of the net strangeness arising from the baryon sector with 
that from the meson sector, which demands the $\mu_S$ to be proportional 
to the $\mu_B$. The variance of $\mu_S$ is within the uncertainties for 
the three values of $K_M$, which indicates that the higher mass strange 
mesons and baryons have negligible influence from the mean-field 
repulsion, while performing the freeze-out parametrization with yields.  

The resulting values of the freeze-out volume (which is presented in the 
freeze-out radius here), are presented in the right bottom panel. A 
similar non-monotonic behavior with the collision energy was earlier 
observed from the chemical freeze-out parametrization with ideal HRG in 
Refs.~\cite{Andronic:2005yp, Chatterjee:2015fua}. The interesting 
observation here is the higher value of freeze-out radius while we imply 
a higher value of $K_M$. A higher value of $K_M$ and $K_B$ suppresses the 
number density, which in turn results in a higher value of freeze-out 
radius to fit the yields. One can see that among the above-discussed 
parameters, the variation of volume with $K$ is much more prominent. It 
seems that considering the repulsive interaction affects the value of the 
freeze-out volume mostly as the yield is directly proportional to the 
volume. 

\begin{figure*}[hbt]
\hspace*{-0.9cm}
\includegraphics[height=7.4 in,width=7.4 in]{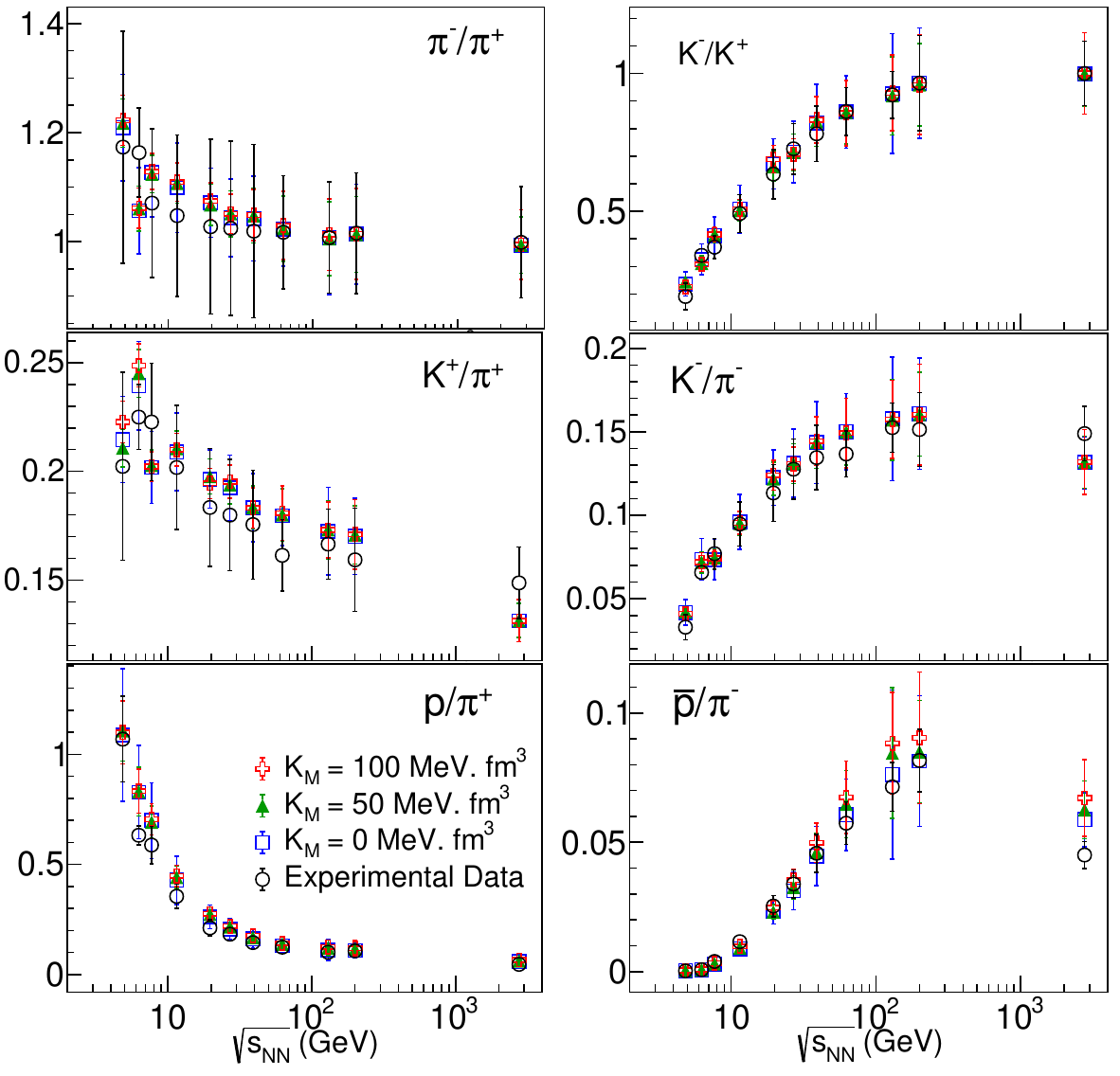}
\caption{$\sqrt{s_{NN}}$ variation of the particle to particle ratios 
$\pi^-/\pi^+$ and $K^-/K^+$ (top), strange to non-strange meson, $K^+/\pi^+$ 
and $K^-/\pi^-$ (middle), non-strange baryon to meson, proton to pion (bottom).
Blue, green, and red points are the results for $K_M$ values of 0, 50, and 
100 $\text{MeV fm}^{-3}$ respectively. Black points are the experimental 
data from Ref.~\cite{STAR:2021iop}.} \label{fig:ratio}
\end{figure*}

\begin{figure}
\hspace{-15mm}
\includegraphics[height=3.2 in,width=3.9 in]{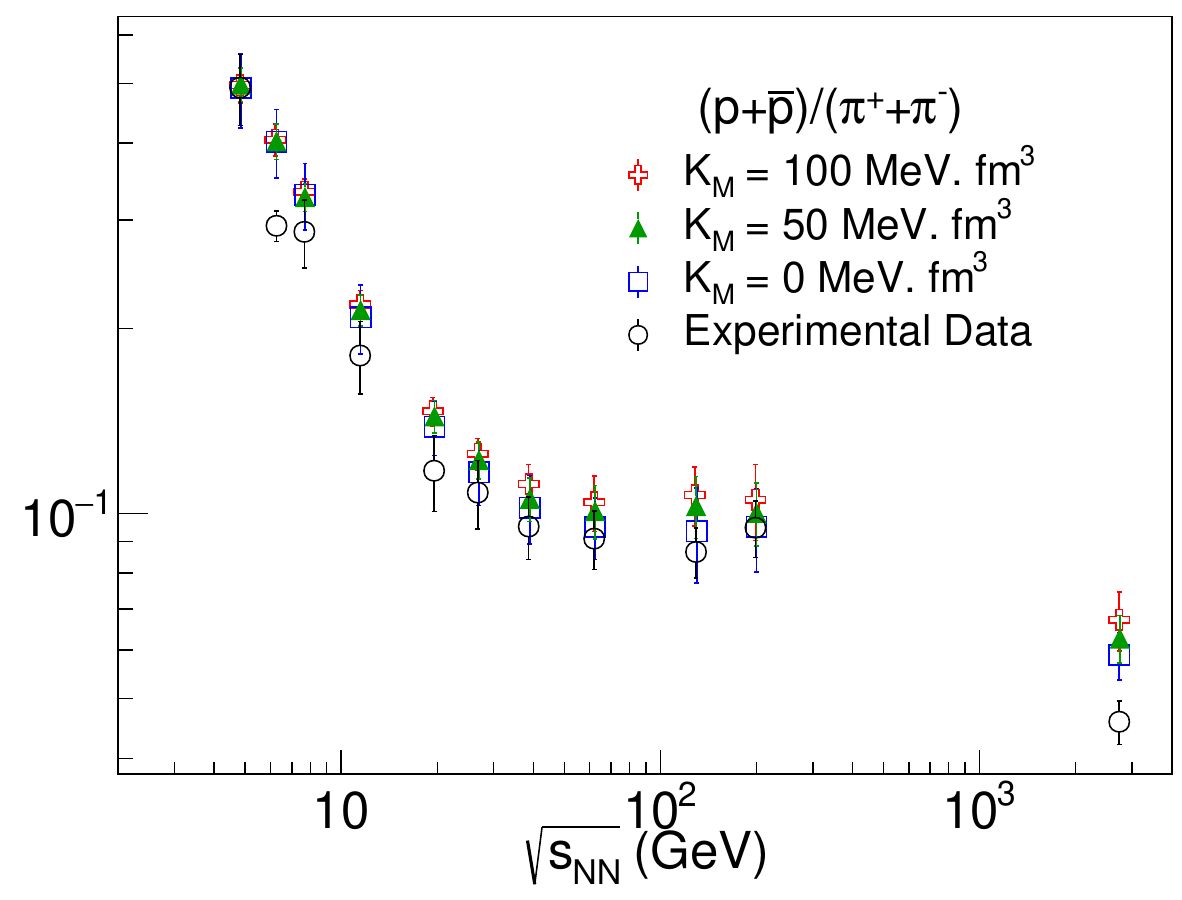}
\caption{$\sqrt{s_{NN}}$ variation of the total proton to total pion. 
Blue, green, and red points are the results for $K_M$ values of 0, 50, 
and 100 $\text{MeV fm}^{-3}$ respectively. Black points are the 
experimental data from Ref.~\cite{STAR:2021iop}.} \label{fig:ppisum}
\end{figure}
\subsection{Particle yields from thermal parametrization:}
To examine the differences in thermal abundances resulting from different 
values of $K_M$, it would be informative to analyze the variations in 
yields. Fig.~\ref{fig:yield} displays the number density of pions, 
kaons, protons, and lambdas, calculated with the resulting 
parametrization. The impact of varying $K_M$ is more pronounced for 
lighter mass pions, while the effect diminishes as the particle mass 
increases. Baryons with higher masses show negligible variations across 
the three cases, whereas pions demonstrate more significant alterations 
when different $K_M$ values are considered.

The effective chemical potential $\mu_{\text{eff}, i} = \mu_i - K 
n_{M, B, \bar{B}}$ is expected to have a significant impact on pions 
since they carry only electric charge, and the magnitude of $\mu_Q$ is 
much smaller compared to other chemical potentials. Conversely, the 
effect of this shift diminishes for strange and non-strange baryons, as 
their respective chemical potentials have larger magnitudes. It is worth 
noting that the chemical potentials themselves are modified for different 
$K_M$ values, contributing to the observed variations. 

In this context, it is important to consider the decay feed-down effect 
as well. The total pion density receives a significant contribution from 
the decay of higher-mass meson and baryon resonances. The suppression of 
these states is also reflected in the final pion abundance, resulting in 
substantial variations. This effect is similarly observed for the 
lowest-mass strange hadron, kaon. On the other hand, baryons receive 
contributions from higher-mass baryons that are already thermally 
suppressed, leading to insignificant variations while considering 
different $K_M$ values. 

At this juncture, we want to reiterate that the yield $dN/dY$ is a product 
of this thermal density and the freeze-out volume $dV/dY$. A reverse trend 
was observed for the freeze-out volume in Fig.~\ref{fig:parameters}, i.e. 
a higher value of $K_M$ resulted in higher values of freeze-out volume. 
The cumulative effect of these two ensures the agreement between the yield 
data and our thermal model estimation. This indicates that the resulting 
parameters (especially freeze-out volume and $K_B$) are dependent on each 
other and on the values of $K_M$. In our present study, it becomes 
challenging to decouple this systemic dependency.

\subsection{Particle ratios:}
It would be interesting to estimate various particle ratios and compare 
them with those from the experimental data. Along with checking the 
efficacy of our parameterization, this will also examine the effect of 
various choices of $K_M$ on thermal yields. Here we shall discuss some of 
the important particle ratios from various sectors.

The ratios of $\pi^-/\pi^+$ and $K^-/K^+$ as a function of $\sqrt{s_{NN}}$ 
are depicted in the upper panel of Fig.~\ref{fig:ratio}. Our 
parametrization successfully reproduces the observed variation of the 
experimental data. The pion ratio is greater than unity at lower 
$\sqrt{s_{NN}}$ due to the higher abundance of neutrons in the colliding 
nuclei, which induces an isospin asymmetry favoring $\pi^-$. However, this 
asymmetry diminishes at higher RHIC and LHC energies, resulting in similar 
yields of $\pi^-$ and $\pi^+$. In the case of kaons, the variation follows 
the trend of $\mu_S$. At lower collision energies, the positively charged 
kaon ($K^+$) becomes more abundant than the negatively charged kaon 
($K^-$) to maintain strangeness neutrality. As the collision energy 
increases, this effect disappears, and the yields of particles and 
antiparticles become equal at the LHC. The qualitative behavior is the 
same for all three values of $K_M$. It seems that the effect of $K_M$ does 
not result in the large variation of the mentioned ratio.

In the context of the heavy-ion collision, the strange to non-strange 
ratios signify the relative abundance of strangeness and portray the 
degree of equilibration for the strange sector \cite{ALICE:2017jyt}. 
Deviations from the equilibrium values have earlier been observed for non-
central collisions, which necessitates the use of a strangeness saturation 
factor $\gamma_S$ \cite{Biswas:2020dsc}. Being the lightest strange to non-
strange particle, the ratio $K^+/\pi^+$ and $K^-/\pi^-$ are widely studied 
within the thermal model. The explanation of the non-monotonic behavior of 
the $K^+/\pi^+$ was discussed as a signature of the thermalization in the 
strange sector and a possible existence of initial partonic 
state~\cite{Gazdzicki:1998vd, Gazdzicki:2003fj, Bratkovskaya:2004kv, 
Koch:2005pk, Nayak:2010uq}. Although these details are beyond the scope of 
the present thermal model, our parameterization suitably explains the data 
for all three values of $K_M$ in the middle panel of Fig.\ref{fig:ratio}. 
Although, there is not much variation among the estimations from the three 
cases, indicating that these ratios have a weak dependence on the variation of 
$K_M$ and $K_B$.

We have shown the proton-to-pion ratio in the bottom panel of 
Fig.~\ref{fig:ratio}. As we have separately used two different mean-field 
coefficients $K_M$ and $K_B$ for the meson and baryon sectors separately, 
this ratio will portray their effect on the respective variation. We have 
plotted $p/\pi^+$ and $\Bar{p}/\pi^-$ to nullify the effect of the charge 
chemical potential. In the context of heavy-ion collision, the abundance 
of pions is mainly dominated by the temperature as they are the lowest 
mass hadrons, whereas the protons mimic the variation of exponential of 
$\mu_B/T$. At lower collision energy the medium is dominated by the 
baryons due to the baryon stopping, whereas at higher collision energies 
the system is dominated by the mesons, and changes from a baryon-dominated 
freeze-out to meson-dominated freeze-out occurs \cite{Cleymans:2004hj}. 
This phenomenon explains the variation observed in the proton to $\pi^+$ 
ratio. On the other hand, the production of anti-proton increases as the 
collision energy increases and at high RHIC and LHC energies, the two 
ratio becomes similar. Here the values of the $p/\pi^+$ ratio increase as 
we increase the $K_M$ for a given collision energy. A higher value of 
$K_M$ suppresses the abundance of pions and produces a higher value of the 
ratio.  

To quantify the impact of the various choice of repulsive parameters $K_M$ 
and $K_B$ in the particle ratios, we have plotted the total proton 
($p + \Bar{p}$) abundance normalized to total pion ($\pi^{+}+\pi^{-}$) 
in Fig.~\ref{fig:ppisum}. This ratio has a larger impact from various 
choices of $K_M$ than that of the individual ratios, as it signifies the 
relative abundance of the lowest mass baryons to that of the lowest mass 
mesons. The parametrization for $K_M=0$ seems to agree with the data 
better than the other choices. At the freeze-out parametrization, one 
should not expect much variation in the baryon yield from the variation of 
$K_B$ due to their heavier masses, on the contrary, the pion yields get 
significantly suppressed for a higher value of $K_M$ due to the lower 
masses. This results in the variation shown as the ratio 
($(p + \bar{p})/(\pi^{+}+\pi^{-})$), as for a given collision energy it 
increases for a higher value of $K_M$. The difference is much more 
pronounced at higher collision energies, as the thermal medium is meson
dominated, so the different choices of $K_M$ produce a larger effect.


\begin{figure}
\hspace*{-0.7cm}
\includegraphics[height=7 in,width=3.8 in]{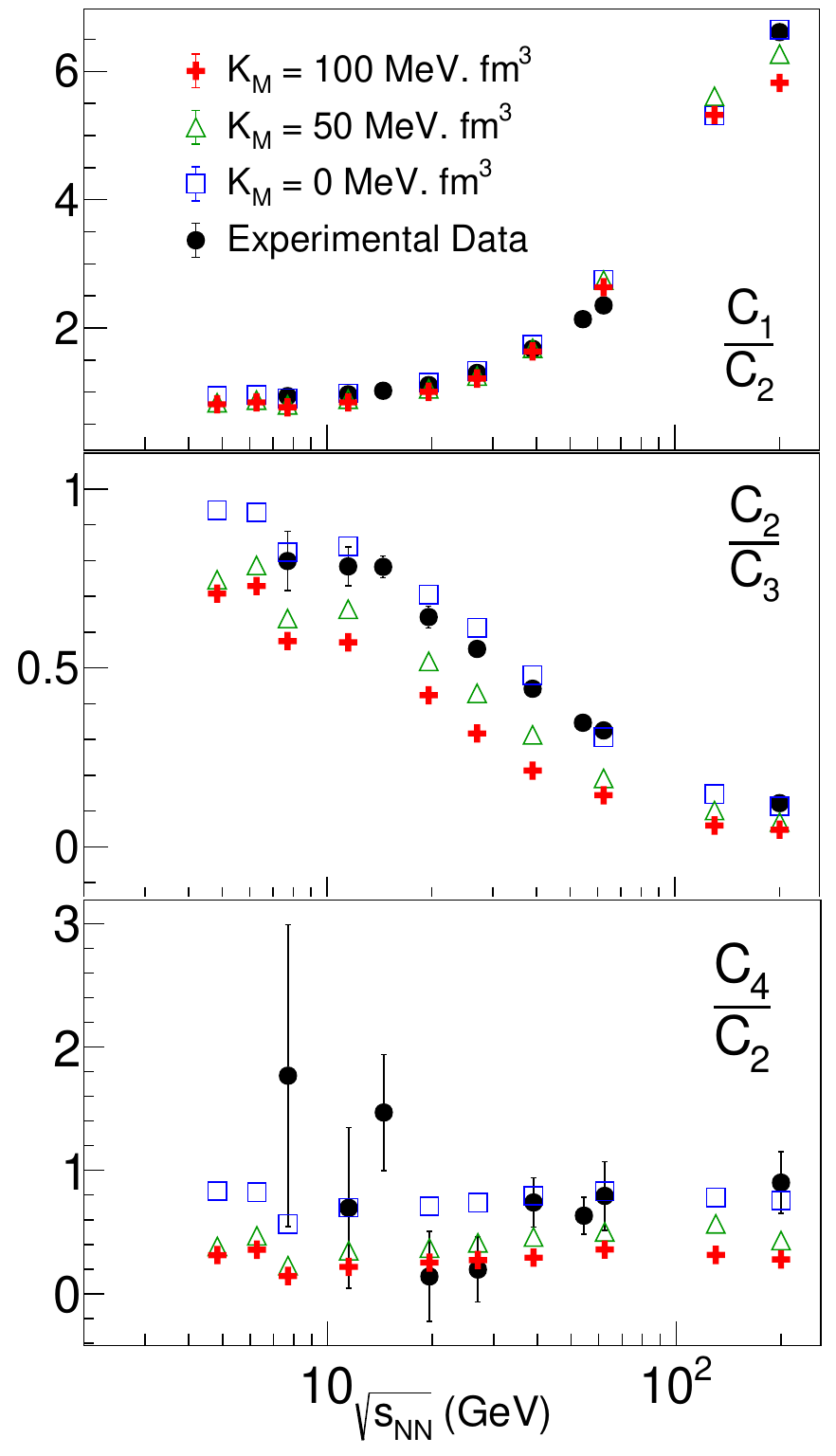}
\caption{$\sqrt{s_{NN}}$ variation of the cumulant ratios $C_2/C_1$ 
(top), $C_3/C_2$ (middle) and $C_4/C_2$ (bottom). Blue, green, and red 
points are the results for $K_M$ values of 0, 50, and 100 
$\text{MeV fm}^{-3}$ respectively. Black points are the experimental data
from Ref.~\cite{STAR:2021iop}.}
\label{fig:chi}
\end{figure}

Motivated by the fact that the ratio corresponding to the total proton 
yield to pions gets significant variation from the values of mean-field 
parameter $K_M$, we have investigated their impact on the ratios of 
susceptibilities calculated with the freeze-out parameterization. The $n$th order susceptibility is defined as:
\begin{equation}
\chi_{x}^{n}=\frac{1}{V T^{3}} \frac{\partial^{n}(\ln Z)}{\partial\left(\frac{\mu_{x}}{T}\right)^{n}}
\end{equation} 
where $\mu_x$ is the chemical potential for conserved charge $x$. The 
susceptibilities would be related to the cumulants measured in the 
heavy-ion collisions as:
\begin{equation}
V T^{3}\chi_{x}^{n}= C_n .
\end{equation}
As we have fixed the $K_M$ and fitted the mean-field co-efficient $K_B$, 
it will be interesting to check the variance in the baryon cumulant 
ratios. We have calculated these cumulants within the Boltzmann 
approximation as it provides a reasonable baseline for 
the massive hadrons and resonances (except $\pi$) along the chemical 
freeze-out boundary \cite{Wheaton:2004qb}, as $m_i-\mu_i >> T$ at the 
respective freeze-out parametrization. Within this consideration, we can 
approximate the interacting partition functions in the Boltzmann limit and 
calculate the $\chi_B^n$ \cite{Huovinen:2017ogf, Li:2022lxk}. The 
differences arising from various values of the $K_M$ increase as we move 
to ratios of higher-order cumulants. The effect is negligible for 
$C_2/C_1$, while $C_3/C_2$ and $C_4/C_2$ decrease as we fix the $K_M$ to 
higher values. As we imply a higher value of $K_M$, it produces a higher 
value of $K_B$ as discussed earlier in Sec.~\ref{Sec:parameters}, which 
translates into these differences. We want to mention that the ratio 
$C_4/C_2$ is $1$ at all collision energy for the ideal HRG case, whereas 
the impact of interaction gives rise to the observed variation. 

As a baseline, we have also plotted results for the net proton cumulants 
estimations from STAR collaboration \cite{STAR:2020tga, STAR:2021iop}. For 
simplification, we have not mimicked the experimental specification like 
$p_T$ cut, decay feed-down into the cumulants calculations. Although the 
effect of decay feed-down and $p_T$ cut-off have been found to be minimal 
earlier \cite{Garg:2013ata, Mishra:2016qyj, Li:2022lxk}. We want to 
reiterate that we have calculated the baryon cumulants ratios, which is 
different from the net-proton ratios. Although the qualitative behavior is 
similar, the quantitative difference between these two increases for higher 
order cumulant ratios \cite{STAR:2021iop}. The non-monotonic variation of 
$C_4/C_2$ is not well captured in the thermal model estimation, although 
the results vary from the ideal baseline of $1$. The $C_2/C_1$ and 
$C_3/C_2$ estimations agree with the data for 
$K_M=50~\text{MeV fm}^{-3}$, there are larger deviation for $C_4/C_2$, 
which seem to match for higher values of $K_M$. This behavior suggests that 
a complete study of the net-proton cumulants with experimental constraints 
might restrict the variation of $K_M$ and $K_B$ both.



\section{Summary}\label{sec:summary}
Recent advancements in incorporating repulsive interactions between 
baryons and mesons in the hadron resonance gas (HRG) model have 
established it as a suitable candidate for providing a bulk description of 
the QCD medium below the transition temperature. Phenomenological 
descriptions such as the excluded volume HRG and van der Waals HRG models 
consider parameters such as a hard-core impenetrable radius of the 
hadrons. On the other hand, the mean-field repulsive HRG model (MFHRG) 
provides a robust representation of the medium by accounting for a density-
dependent interaction strength. However, this model requires the inclusion 
of parameters such as $K_B$ and $K_M$ to scale the interaction strength 
among baryons and mesons, which can be appropriately estimated using bulk 
observables obtained from lattice QCD \cite{Huovinen:2017ogf, 
Kadam:2019peo}. It is crucial to apply this mean-field repulsive model to 
analyze data from heavy-ion collision experiments and assess its 
effectiveness in comparison to other counterparts such as the ideal HRG, 
evHRG, vdWHRG, and so on. Exploring the chemical freeze-out surface 
provides a foundation for investigating the collision energy dependence of 
the repulsive interaction strength by estimating $K_M$ and $K_B$.

To parametrize the chemical freeze-out surface, we utilized the $p_T$-
integrated mid-rapidity yield $dN/dY$ data for pions, kaons, protons, 
$\Lambda$, and $\Xi$ in the most central collisions. The collision energy 
range available in AGS (4.85 GeV), RHIC-BES, and LHC (2.76 TeV) was 
analyzed. Given that the parameters $K_B$ and $K_M$ are interdependent due 
to relevant constraints and decay feed-down effects, evaluating them 
independently can lead to larger numerical variances. To address this 
issue, we fixed $K_M$ at three representative values (0, 50, and 100 
MeV.fm$^{-3}$) and performed a $\chi^2$ fitting to determine the remaining 
parameters: $T$, $\mu_B$, $\mu_Q$, $\mu_S$, $K_B$, and the freeze-out 
radius $R$.

While the values of $K_B$ were found to be finite and influenced the 
goodness of fit, the other parameters were consistent with those obtained 
from the ideal HRG model. Notably, $K_B$ increases with collision energy 
and becomes significantly higher at higher $\sqrt{s_{NN}}$. It is 
intriguing to observe that the values of $K_B$ obtained from this freeze-
out analysis are similar to those from earlier studies using the mean-
field approach. The agreement between the estimation of $K_B$ from lattice 
QCD-motivated studies \cite{Huovinen:2017ogf, Huovinen:2018xmq, 
Kadam:2019peo, Pal:2021qav} and our analysis underscores the effectiveness 
of this model in describing the bulk properties of the created medium in 
heavy-ion collisions. 

Studying the influence of repulsive interactions on the thermal abundance 
of different states was crucial. While the effect of finite $K_M$ and 
$K_B$ values on the number density of massive strange hadrons and baryons 
was not significant, it played a more prominent role in the case of pions. 
Particle ratios within the same sector, such as meson-to-meson and baryon-
to-baryon ratios were less affected by variations in $K_M$ and $K_B$. 
However, the proton-to-pion ratios exhibited significant variations. 
Consequently, the total proton to total pion ratio became a subject of 
investigation, as it appeared to be strongly dependent on the values of 
$K_M$. Additionally, we explored the ratio of baryon susceptibilities 
using this freeze-out parameterization, as these susceptibilities are 
linked to net-proton cumulants measured in heavy-ion collisions. While our 
freeze-out parametrization was based on yields, there was a general 
agreement between our estimations of baryon cumulant ratios and the 
measurements of net-proton for lower orders. However, discrepancies 
arose when considering fourth-order cumulants. Proper treatment of 
cumulant ratios requires accounting for decay feed-down effects and 
implementing $p_T$ cuts within the framework of this mean-field repulsive 
HRG model. This consideration will be essential for future studies, 
particularly in the context of energy available in BES-II, HADES, and CBM 
experiments.

\section*{Acknowledgements}
D.B expresses gratitude to Sayantan Sharma, Aman Kanojia, Somenath Pal and 
Hiranmaya Mishra for engaging and fruitful discussions. D.B. would like to 
express sincere gratitude for the support received from NISER, 
Bhubaneswar, with special thanks to A. Jaiswal for the kind assistance and 
hospitality during the visit, where the majority of this work was 
performed.
\appendix
\begin{table*}[hbt]
	\caption{ }
	\begin{center}
		\begin{tabular}{|c|c|c|c|c|c|c|c|}
			\hline
			\multicolumn{8}{|c|}{Particle+Anti Particle: $\pi,K,P,\Lambda,\Xi$, centrality: 0-5\%, Constrain: $netB/netQ=2.5$ and $netS=0.0$ } \\
			\hline
			$\sqrt{s_{NN}}$ (GeV) & $\kappa_M $ (MeV.fm$^{-3}$) & $\kappa_B$ (GeV.fm$^{-3}$) &$T_{ch} \ \text{(MeV)}$ &$\mu_B \ \text{(MeV)}$ &  $\mu_S \ \text{(MeV)}$ &R(fm)&$\chi^2/ndf$\\

			\hline
			
			&0&0.03(0.05)&124.50(3.70)&526.02(21.50)&112.02(9.39)&6.25(0.26)&1.11\\ \cline{2-8}   
			&50&0.16(0.05)&125.08(3.73)&532.34(21.39)&111.11(10.28)&6.68(0.27)&1.72\\ \cline{2-8}   
			\multirow{-3}{*}{4.83} 
			&100&0.18(0.0)&125.85(3.38)&540.40(22.24)&117.59(11.32)&6.80(0.29)&1.67\\ \hline
			&0&0.04(0.04)&131.03(2.95)&454.88(14.25)&93.27(14.20)&6.77(0.37)&2.60\\ \cline{2-8}   
			&50&0.18(0.06)&130.57(3.13)&450.07(14.91)&96.07(13.50)&6.96(0.38)&2.81\\ \cline{2-8}   
			\multirow{-3}{*}{6.27}			&100&0.22(0.05)&131.43(2.95)&463.34(17.58)&98.04(14.04)&7.16(0.37)&2.65\\ \hline
			&0&0.07(0.06)&147.50(3.40)&428.35(11.53)&87.62(10.38)&5.40(0.38)&2.06\\ \cline{2-8}   
			&50&0.19(0.08)&148.87(3.85)&429.77(12.07)&88.05(10.40)&5.57(0.40)&2.06\\ \cline{2-8}   
			\multirow{-3}{*}{7.7} 
			&100&0.25(0.09)&149.26(4.41)&431.36(13.29)&88.93(10.60)&5.71(0.44)&2.06\\ \hline
			
		& 0&0.07(0.08)& 153.01(3.13)&317.27(12.37)&68.51(8.30)&5.60(0.30)&1.56\\ \cline{2-8}  
			&50&0.22(0.10)&153.85(3.33)&319.68(15.63)&70.94(9.33)&5.84(0.34)&1.57\\ \cline{2-8}  
			\multirow{-3}{*}{11.5}  &100&0.33(0.10)&154.55(4.60)&320.75(14.59)&71.05(9.45)&6.05(0.36)&1.54\\ 	\hline
			&0&0.10(0.06)&160.03(3.56)&207.65(12.45)&44.15(7.33)&5.56(0.32)&1.18\\ \cline{2-8}  
			&50&0.32(0.12)&160.93(4.55)&212.82(12.51)&45.75(9.34)&5.77(0.35)&1.19\\ \cline{2-8}  
			\multirow{-3}{*}{19.6}
            &100&0.48(0.18)
			&161.46(4.44)&213.43(09.45)&43.08(7.29)&6.07(0.35)&1.20\\ \hline
			&0&0.12(0.07)&160.10(3.42)&162.02(14.70)&35.70(9.77)&5.76(0.28)&1.07\\ \cline{2-8}  
			&50&0.37(0.12)&162.27(3.02)&165.41(10.69)&37.52(10.26)&5.89(0.29)&0.99\\ \cline{2-8}  
			\multirow{-3}{*}{27.0} 
            &100&0.58(0.16)&
			163.77(4.43)&166.64(10.39)&39.08(8.43)&6.13(0.31)&1.09\\ \hline
		    &0&0.12(0.07)&160.81(4.64)&112.91(10.75)&23.89(10.56)&5.85(0.28)&0.54\\ \cline{2-8}  
			&50&0.42(0.12)&163.24(5.75)&113.83(08.94)&24.00(8.74)&6.01(0.30)& 0.47\\ \cline{2-8}  
			\multirow{-3}{*}{39.0} 
            &100&0.65(0.18)&165.98(4.53)&116.71(08.57)&24.28(8.62)&6.25(0.34)&0.43\\ \hline
            &0&0.10(0.05)&164.89(4.50)&69.91(8.68)&18.04(7.25)&6.13(0.30)&1.34\\ \cline{2-8}  
			&50&0.38(0.09)&166.80(4.90)&70.21(7.53)&18.79(9.87)&6.44(0.32)&1.23\\ \cline{2-8}  
			
		\multirow{-3}{*}{62.4}&100&0.59(0.15)&
		167.11(5.23)&73.07(7.29)&19.87(10.90)&6.65(0.34)&1.17\\ \hline
			 &0&0.14(0.08)&166.70(5.46)&35.79(12.85)&9.70(8.00)&6.16(0.34)&1.62\\ \cline{2-8}  
			&50&0.28(0.10)&170.48(5.83)&35.27(8.48)&10.93(7.88)&6.61(0.37)&1.06\\ \cline{2-8}  
			\multirow{-3}{*}{130.0}&100&0.55(0.10)&
			172.80(5.33)&37.39(6.08)&11.12(5.80)&6.71(0.38)&1.16\\ \hline
			
		&0&0.14(0.07)& 169.57(4.59)&28.59(7.61)&6.91(7.61)&6.81(0.40)&1.03\\ \cline{2-8}  
			&50&0.42(0.13)&171.17(4.67)&30.72(7.73)&7.76(6.70)&6.90(0.40)&0.69\\ \cline{2-8}  
			
			\multirow{-3}{*}{200.0} & 100&0.61(0.24)&173.15(5.22)&33.34(8.74)&8.20(6.97)&7.13(0.42)&0.57
			\\ \hline
			
			 & 0&0.20(0.11)& 158.03(5.62)&1.97(1.30)&0.0&10.75(0.46)&2.54\\ \cline{2-8}  
			&50&0.45(0.18)&158.77(5.03)&1.11(0.54)&0.0&10.95(0.46)&2.31\\ \cline{2-8}  
			
			\multirow{-3}{*}{2760.00} &
			100&0.66(0.20)&159.85(5.74)&1.06(0.74)&0.0&11.15(0.45)&2.43
			\\ \hline
			
		\end{tabular}
\end{center}\label{tb:1}
\end{table*}

\bibliographystyle{elsarticle-num}
\bibliography{referencesMF.bib}

\end{document}